\documentstyle[twocolumn,floats,ifthen,aps,prb,psfig]{revtex}

\newcommand\eps{\varepsilon}

\newcommand\N{N}

\begin{document}

\title{Theory of transient spectroscopy \\
of multiple quantum well structures}
\author{M.~Ershov}
\address{Department of Physics and Astronomy, Georgia State University,
Atlanta, Georgia 30303}
\author{H. Ruda, A. Shik}
\address{Energenius Centre for Advanced Nanotechnology,
University of Toronto,\\
Toronto M5S 3E3, Canada}
\author{A.~G.~U.~Perera}
\address{Department of Physics and Astronomy, Georgia State University,
Atlanta, Georgia 30303} \maketitle

\begin{abstract}
A theory of the transient spectroscopy of quantum well (QW)
structures under a large applied bias  is presented. An
analytical model of the initial part of the transient current is
proposed. The time constant of the transient current depends not
only on the emission rate from the QWs, as is usually assumed,
but also on the subsequent carrier transport across QWs.
Numerical simulation was used to confirm the validity of the
proposed model, and to study the transient current on a larger
time scale. It is shown that the transient current is influenced
by the nonuniform distribution of the electric field and related
effects, which results in a step-like behavior of the current. A
procedure of extraction of the QW emission time from the
transient spectroscopy experiments is suggested.
\end{abstract}




\section{Introduction}
Transient spectroscopy of quantum-well (QW) structures allows to
study the emission processes from the QWs and thus to obtain
information on QW parameters, such as the energy spectrum,
photoionization cross-section, tunneling escape time,
etc.~\cite{MartinQW83,MartinetTS91,LucTS93} This technique is
based on an analysis of the transient current or capacitance
relaxation upon the application of a {\em large-signal} bias
across the QW structure. It complements admittance spectroscopy,
which studies the {\em ac} current in the QW structure upon
application of a {\em small-signal} voltage.~\cite{LangAS87} The
transient spectroscopy of QWs has many similarities with the deep
level transient spectroscopy (DLTS)~\cite{LangDLTS74} and enables
a simple theory to be derived for use in the processing of
experimental data.~\cite{LucTS93} However, there is an important
difference in the carrier capture by deep levels and that
involving QWs. In the latter case, the presence of a continuous
energy spectrum for the in-plane motion in the QW allows the
capture by emission of a single optical phonon rather than by a
multi-phonon processes typical for deep levels.~\cite{APY} As a
result, the corresponding capture times are several orders of
magnitude less than for deep levels and often do not exceed a few
picoseconds.~\cite{BSHW} This quantitative difference results in
serious qualitative consequences. The processes of carrier
transport between neighboring QWs can no longer be considered as
infinitely fast. These processes may play a decisive role in the
relaxation kinetics changing noticeably the formulae of a simple
theory, similar to the case of structures with very high
concentration of deep levels. \cite{ShikCS84} In this work we
present a theoretical description of the transient spectroscopy
of QWs and discuss its possible applications. We obtain more
general analytical expressions for the parameters of the
transient current than those of Ref.~\onlinecite{LucTS93}. The
analytical model is confirmed by numerical simulation, and the
procedure of the extraction of the QW parameters from
experimental data is discussed.

\section{Analytical model}
We consider a QW structure containing $M$ QWs (n-doped at sheet density
$\N_D$) of width $L_{w}$ separated by undoped barriers of width $L_{b}$
large enough to prevent inter-well tunneling (see Fig.~1). This structure
is typical for quantum well infrared photodetectors (QWIPs).~\cite{LevineR}
The QW structure is provided with a heavily doped (Ohmic) collector and a
blocking emitter contact (for example, containing a Schottky barrier or p-n
junction) which is often used to avoid DC current and thus to simplify the
interpretation of experimental data.~\cite{LucTS93}

During the first period of the transient spectroscopy experiment,~\cite
{LucTS93} the forward bias is applied to the emitter, and all QWs are
filled by electrons with the equilibrium sheet density $\N_{0}\approx
\N_{D}$. In the second period, a large reverse bias $V$ is applied to
extract electrons from the QWs, and transient current is recorded. The
problem has some similarities with the treatment presented for the kinetics
of electron packet in a system of undoped QWs.~\cite{Ryvkin}. Immediately
after the application of $V$, the electric field in the QW structure is
uniform and given by $E_{0}=(V+V_{bi})/L$, where $V_{bi}$ is the built-in
voltage between the emitter and collector, and $L=ML_{w}+(M+1)L_{b}\approx
(M+1)L_{b}$ is the structure thickness. This field causes fast removal of
delocalized electrons from the structure at almost fixed $\N_{0}$. This is
a very fast (on a ps time scale) component of the transient current
\cite{ErshovAPLTr} limited by carrier capture and transit times, which is
manifested as an instantaneous current step in the case of limited time
resolution of a measurement setup. We shall be interested here in a
subsequent slow current relaxation caused by the QW recharging. The initial
part of this relaxation (after completion of the fast transient) can be
easily calculated.

In the presence of external illumination, the emission rate from the $i$-th
QW with electron density $\N_{i}$ is $G N_i={\sigma \Phi N_i +\gamma \exp
[(\eps_f-\eps_i)/kT]}$, where $\sigma $ is the photoionization
cross-section, $\Phi $ is the incident photon flux, $\eps_i$ is the QW
ionization energy, $\eps_f(\N_i)$ is the Fermi energy of the electrons in
the QW, and $\gamma $ is the thermoionization coefficient. In our estimates
and further numerical simulations, we shall restrict our analytical
calculations to the case of relatively low temperatures or high light
intensities giving $G\cong \sigma \Phi$. This restriction is not compulsory
since all analytical formulae are applicable for an arbitrary relation
between the optical and thermal generation.

We assume that the carriers emitted from the QWs drift with a constant
velocity $v_{d}$ towards the collector. While traversing a QW, carriers are
captured by the QW with a probability $p$ ($0<p\leq 1$). Carriers emitted
from the $k$-th QW give the following contribution to the carrier
concentration in the $i$-th barrier (between $i$-th and $(i+1)$-th QWs):
\begin{equation}
n_{ki}=\frac{G\N_{k}}{v_{d}}\,(1-p)^{i-k},\qquad (i\geq k). \label{conc}
\end{equation}
The total concentration in the $i$-th barrier $n_{i}$
\begin{equation}
n_{i}=\sum_{k=1}^{i}n_{ki}=\frac{G}{v_{d}}\sum_{k=1}^{i}\N_{k}(1-p)^{i-k}.
\label{conc1}
\end{equation}
Since the change of a QW charge is determined by the balance between carrier
capture and emission, we can, with the help of Eq. (\ref{conc1}), obtain the
system determining the kinetics of all $\N_{i}$:
\begin{equation}
\frac{d\N_{i}}{dt}=-G\N_{i}+n_{i-1}v_{d}\,p=G\left[ -\N_{i}+%
\sum_{k=1}^{i-1}\N_{k}p(1-p)^{i-1-k}\right]  \label{system}
\end{equation}
and, hence, of the current in the external circuit $I(t)$ which could be
expressed in terms of $\N_{i}(t):$
\begin{equation}
I(t)=\frac{ev_{d}}{M+1}\sum_{i=1}^{M}n_{i}=\frac{eG}{M+1}\sum_{i=1}^{M}
\sum_{k=1}^{i}\N_{k}(t)(1-p)^{i-k}.  \label{I(t)}
\end{equation}

In principle, we can obtain analytical (though rather cumbersome) solution
of the linear system of Eq.~(\ref{system}) for an arbitrary $t$. However, it
would not be correct. The change in $\N_{i}$ causes re-distribution of the
electric field and, hence, of the drift velocity in the system. This means
that $v_{d} $ is no longer constant but changes from point to point in an
unknown way so that the behaviour of $I(t)$ remains unknown. That is why we
restrict ourselves to the initial stage of the slow relaxation when we can
still assume that in the right-hand side of Eq. (\ref{system}) $%
\N_{i}=\N_{0}$ and $v_{d}={\rm const}.$ This gives
\begin{eqnarray}
 \frac{d\N_{i}}{dt} &=&-G\N_{0}(1-p)^{i-1};  \label{dN} \\
 I(0)&=&\frac{I_{e}}{p(M+1)}\left\{ 1-\frac{1-p}{pM}\left[
        1-(1-p)^{M}\right]\right\} ;  \label{I(0)} \\
 \frac{dI}{dt}(0) &=&-\frac{G
I_{e}}{p^2M(M+1)}\left[ 1-(1+Mp)(1-p)^{M}\right] \label{dI(0)}
\end{eqnarray}
where $I_{e}=eG\N_{0}M$ is the total emission rate from all QWs. Eqs. (\ref
{I(0)}),(\ref{dI(0)}) give us the relaxation time constant (inverse
normalized slope of the current):
\begin{equation}
\tau =-\left( \frac{dI/dt}{I}\right)
^{-1}=\frac{pM-(1-p)[1-(1-p)^{M}]}{G\left[ 1-(1+pM)(1-p)^{M}\right]}.
\label{tau}
\end{equation}

In the most interesting case, when $p\ll 1$ and $M\gg 1$ (which corresponds
to practical QWIPs), parameters $I_{0}$ and $\tau $ are expressed as:
\begin{eqnarray}
I_{0} &=&I_{e}\times g[1-g(1-\exp (-1/g))],  \label{I0g} \\ \tau &=&
\frac{1}{G}\times \frac{1-g[1-\exp (-1/g)]}{g[1-(1+1/g)\exp (-1/g)]},
\label{taug}
\end{eqnarray}

\noindent where $g=1/(pM)$ is a transport parameter. If we characterize QW
capture processes by the capture time $\tau _{c}$ or the capture velocity
$v_{c}=L_{p}/\tau _{c}$, which is related to capture probability as
$p=1/(1+v_d/v_c)$,~\cite {RosQE94,ShadrinN1} then $g=\tau _{c}/\tau
_{tr}+1/M\approx \tau _{c}/\tau _{tr}$, where $\tau _{tr}=L/v_{d}$ is the
transit time.  Therefore, the parameter $g$ corresponds to the photocurrent
gain of a QWIP.\cite{LevineR}

It should be noted that the amplitude of the transient current $I_{0}$ is
equal to the amplitude of the fast transient (primary photocurrent) in a
photoexcited QWIP.~\cite{ErshovAPLTr} In general, the time constant
for the transient current $\tau $ is determined not only by the emission time $%
1/G$, but also by a transport parameter $g,$ similarly to the case of DLTS
for a very high concentration of deep centers.\cite{ShikCS84} Particularly,
in the case $g\ll 1$ we have $\tau \approx 1/(gG)\gg 1/G$. Hence, one cannot
obtain the photoionization cross-section from the transient spectroscopy
experiment ignoring the correction factor (see
Eqs.~(\ref{tau})--(\ref{taug})) dependent on $g$. Only in the limiting case
$g\gg 1$ (or $pM\ll 1$), the relaxation time tends to $1/G$ which
corresponds to the simple model of Ref.~\onlinecite{LucTS93}. To fulfill
this condition, one has to use QW structure with small capture probability
and small number of QWs. In this case $I_{0}=0.5\times I_{e}$, which
corresponds to a high-frequency gain value of 0.5 for extrinsic
photoconductors and QWIPs with large value of the low-frequency gain
$g$.~\cite{ErshovAPLTr,Kaufman69,Blouke72} Note that the capture probability
$p$ is a function of the electric field, decreasing with field, so that the
simplified approach predicting $\tau =1/G$ can be accurate at high fields,
but inaccurate at low fields.

We point out that the value of the photocurrent gain $g$ can be determined
from the transient photocurrent in QWIP illuminated by a step-like infrared
radiation, where the ratio of the amplitude of the fast transient to the
steady-state photocurrent is equal to $\{1-g[1-\exp (-1/g)]\}$.~\cite
{ErshovAPLTr}

\section{Numerical simulation}
The model presented above is justified only for the initial part of the
transient, since we neglected the modulation of the electric field due to
the QW recharging. To check this model and to obtain a description of the
transient current for a wider time interval, we also studied the transient
processes using numerical simulation. A time dependent QWIP simulator~\cite
{ErshovAPLTr,ErshovAPL95} was used with a zero-current boundary condition
for the reverse-biased emitter contact. We simulated the transient
spectroscopy experiment of Ref.~\onlinecite{LucTS93} on GaAs/Al$_{0.25}$Ga$%
_{0.75}$As QW structure with the area $S=2\times 10^{-4}$~cm$^{-2}$
containing 10 donor-doped QWs with $L_{w}=$60~\AA\ QWs and $\N_{D}=5\times
10^{11}$~cm$^{-2}$, undoped barriers with $L_{b}=$350~\AA , Schottky emitter
contact ($V_{bi}=0.75$~V) and collector GaAs contact doped with donors at 10$%
^{18}$~cm$^{-3}.$ The photoexcitation conditions were similar to those used
in the experiment.~\cite{LucTS93}

Figure~2(a) shows the transient current calculated for a reverse bias of
1~V, which for the given $V_{bi}$ corresponds to the applied field $%
E_{0}\approx $40~kV/cm. The capture probability was chosen to be $p=0.04$ so
that $g=2.5$. The initial part of the transient ($t\lesssim \tau ^{\prime }$
where $\tau ^{\prime }\approx 43$ $\mu $s is the position of the first step
in Fig.~2(a)) is very well described by an exponential function with the
amplitude and time constant calculated by our analytical model without any
fitting parameters. The time constant obtained is $\tau \approx 345$~$\mu $%
s, while the emission time is somewhat smaller, $1/G=300$~$\mu $s.

Starting from the time moment $t=\tau ^{\prime }$, the transient current
decays more rapidly, and displays a series of steps and shoulders. These
features are due to the redistribution of the electric field caused by the
depleting of QWs (see Fig.~2(b,c)). The $i$-th step occurs when the electric
field in the $(M+1-i)$-th barrier becomes zero. When this happens, the
electron density in the $(M+1-i)$-th QW returns to its equilibrium value $%
\N_{0}$, and this well does not contribute to the emission current. The
electron transport in the region between this QW and collector is purely
diffusive. Using Eq.~(\ref{dN}) and the condition of zero electric field in
the $M$-th barrier $E_{0}=\sum_{i=1}^{M}\frac{e\Delta \N_{i}}{\varepsilon
\varepsilon _{0}}\frac{i}{M+1}$ ($\varepsilon \varepsilon _{0}$ is the
dielectric constant), we obtain the following estimate for the time constant
$\tau ^{\prime }$:
\begin{equation}
\tau ^{\prime }\approx \frac{\varepsilon \varepsilon _{0}E_{0}}{e\,G\N_{0}M}\,%
\frac{1}{g^{2}\,[1-(1+1/g)\exp (-1/g)]}.  \label{taup}
\end{equation}

\noindent For the case of Fig.~2 this estimate gives $\tau ^{\prime }\approx
39$~$\mu $s, which is in a good agreement with the results of numerical
calculations ($\tau ^{\prime }\approx 43$~$\mu $s).

Since $\tau ^{\prime }\ll \tau $ (unless $g$ is very small), only a small
initial part of the transient process is described by the exponential
function with time constant $\tau $ and amplitude $I_{0}$. Thus, the fitting
of experimental data by an exponential function to extract the time constant
$\tau $ should be done over the interval $0\lesssim t<\tau ^{\prime }$. The
fitting over longer time intervals can result in a significant error in
estimating $\tau $ (the dashed line on Fig.~2(a) is an ``intuitive''
exponential fitting with the time constant $\tau =130$~$\mu $s). It should
be stressed that the measurement circuit should have $RC$-time constant ($R$
is the load resistance and $C$ is the QW structure high-frequency
capacitance) much smaller than time constant $\tau ^{\prime }$ for correct
evaluation of $\tau $.

To check the influence of the QW capture velocity $v_{c}$ on the time
constant $\tau $, we simulated the transient response for different values
of $v_{c}$. The values of the amplitude and time constant of the initial
part of the transient current extracted from numerical simulation are shown
in Fig.~3. A good agreement between these results and formulas
Eq.~(\ref{I(0)}),(\ref{tau}) proves the validity of the analytical model.

So far we have assumed that the photoionization cross-section (or emission
rate) is independent of the local electric field. While this is a good
approximation for bound-to-continuum transitions, the photoionization cross
section for bound-to-bound and bound-to-quasi-bound transitions depends
strongly on electric field.~\cite{LevineR} To investigate this effect, we
compared the results of simulation for the cases of field-dependent and
field-independent cross-section (see Fig.~4). We used the model of the
field-dependent cross-section proposed in Ref.~\onlinecite{LucTS93}:
\begin{equation}
\sigma (E)=\sigma _{0}\times 0.5\text{\thinspace erfc}[(\varepsilon
_{2}-eEL_{w}/2)/(\sqrt{2}\Delta \varepsilon )],  \label{cross}
\end{equation}

\noindent where $\sigma _{0\text{ }}$is the optical cross-section, erfc($x$)
is the complementary error function, $\varepsilon _{2}$ is the ionization
energy of the second level, and $\Delta \varepsilon $ is the variance of $%
\varepsilon _{2}$ due to fluctuations.~\cite{LucTS93} The values $%
\varepsilon _{2}=4$~meV and $\Delta \varepsilon =3.5$~meV also taken from
Ref.~\onlinecite{LucTS93}, were used. It is seen from Fig.~4 that the field
dependence of cross-section washes out steps on the $I(t)$ curve. Moreover,
the initial part of the curve, $0<t<\tau ^{\prime },$ deviates significantly
from an exponential function of the analytical model. Physically, this is
caused by the decrease of the photoemission current from near-collector QWs
due to the electric field redistribution. This effect makes the extraction
procedure of the time constant $\tau $ from the slope of the transient
current more complicated. However, the transient current amplitude $I_{0}$
is not affected by the field redistribution. Since the amplitude is directly
related to the photoemission current ($I_{0}=0.5 I_e$) in the case $g\gg 1$,
we propose to use the amplitude of the transient current rather than its
slope to extract the photoionization cross-section from experimental data.

\section{Conclusions}
A theory of the transient spectroscopy of QW structures is presented.
Analytical expressions for the initial stage of relaxation current are
derived. It is shown that the time constant of the transient current is a
function of both the photoionization cross-section and the transport
parameter $g$ becoming $g$-independent at $g\gg 1$. Numerical simulation is
used to check the validity of the analytical model and study the transient
current in more detail. The procedure of extraction the QW emission rate
from the experimental data is discussed.

This work is supported in part by the NSF under Grant No.~ECS-9809746. H.~R.
and A.~S. also gratefully acknowledge the support of NSERC.


\begin{figure}[bht]
\centerline{\psfig{figure=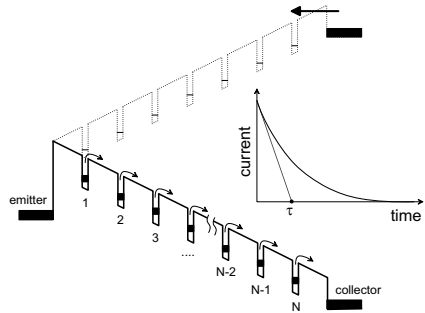,width=8cm}} \vspace{-4cm}
\caption{Schematic illustration of the conduction band diagram and the
transient current in the transient spectroscopy experiment.} \label{fig:1}
\end{figure}

\begin{figure}[tbp]
\centerline{\psfig{figure=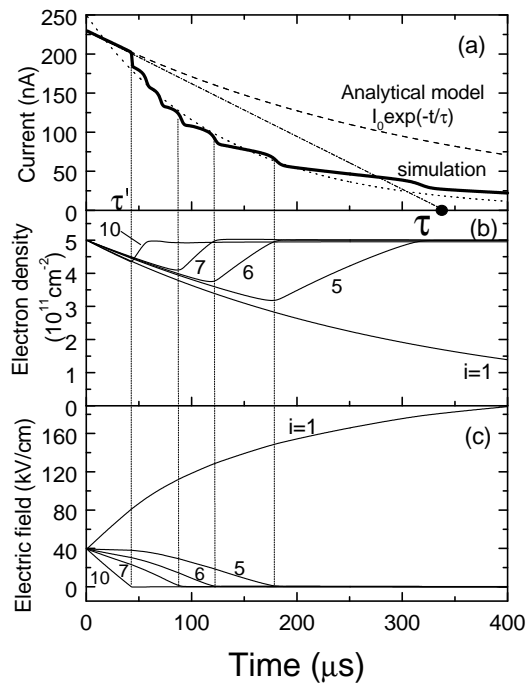,width=8cm}} \caption{(a) Transient
current, (b) electron density in QWs, and (c) electric field in the
barriers for structure with 10 QWs. In (a), solid line -- numerical
simulation, dashed line - analytical model, and dotted line -- exponential
fitting of the transient current on a large time scale. In (b) and (c),
labels indicate the index of the QW and barrier, respectively.}
\label{fig:2}
\end{figure}

\begin{figure}[tbp]
\centerline{\psfig{figure=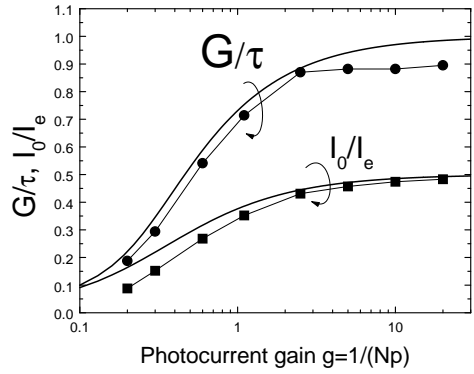,width=8cm}} \caption{Dependence of
the inverse normalized time constant $G/\tau$ and amplitude of the
transient current $I_0/I_e$ on the value of gain $g$. Solid lines --
analytical model, markers -- numerical simulation.} \label{fig:3}
\end{figure}

\begin{figure}[tbp]
\centerline{\psfig{figure=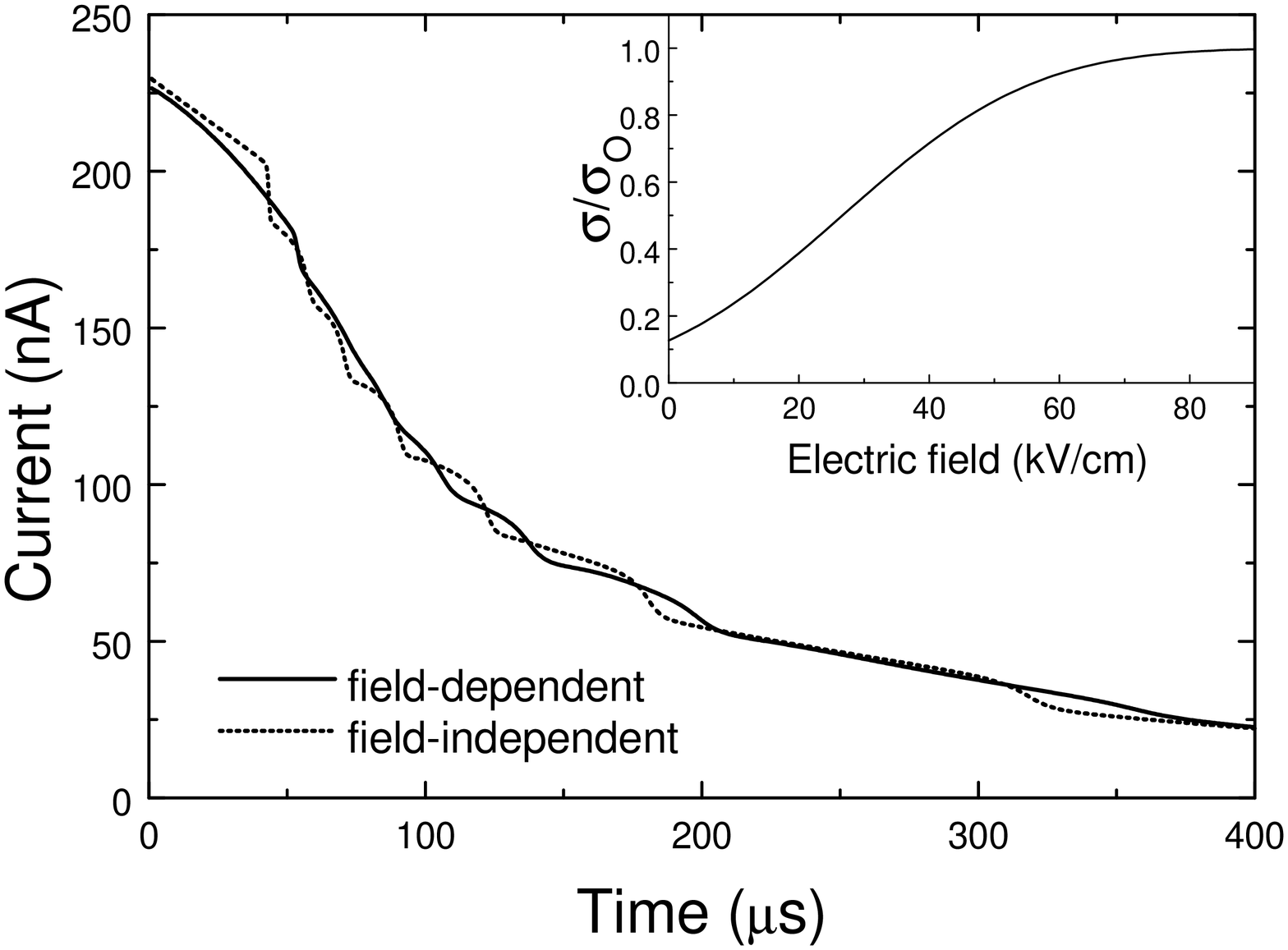,width=8cm}} \caption{Comparison of
the transient currents for the cases of the field-dependent (solid line)
and field-independent (dashed line) photoionization cross-sections. The
inset shows the field dependence of the photoionization cross-section.}
\label{fig:4}
\end{figure}

\end{document}